\documentclass[twocolumn,showpacs,preprintnumbers,amsmath,amssymb,groupedaddress,pra]{revtex4}
\newcommand{\ket}[1]{\mbox{$\vert #1 \rangle$}}
\newcommand{\bra}[1]{\mbox{$\langle #1 \vert$}}
\newcommand{\Rb}{\mbox{$^{87}$Rb}}
\usepackage{dcolumn}
\usepackage{bm}
\usepackage{graphicx} 
\usepackage{txfonts}
\begin{document}
\title{Precision measurement of spin-dependent interaction strengths for spin-1 and spin-2 \Rb\ atoms}
\author{Artur Widera}\email{Widera@Uni-Mainz.DE}
\author{Fabrice Gerbier$^{1}$}
\author{Simon F\"olling}
\author{Tatjana Gericke}
\author{Olaf Mandel$^{2}$}
\author{Immanuel Bloch}
\address{Johannes Gutenberg-Universit\"at, Staudingerweg 7,
55099 Mainz, Germany}
\address{$^1$ Laboratoire Kastler Brossel, D\'epartement de Physique de l'ENS, 75 005 Paris, France\footnote{Laboratoire Kastler Brossel is a research unit of Ecole normale sup\'erieure and Universit\'e Paris 6, associated to CNRS.}}
\address{$^2$ Department of Physics, Stanford University, Stanford, CA 94305, USA}
\date{\today}

\begin{abstract}
We report on precision measurements of spin-dependent interaction-strengths in the \Rb\ spin-1 and spin-2 hyperfine ground states. Our method is based on the recent observation of coherence in the collisionally driven spin-dynamics of ultracold atom pairs trapped in optical lattices. Analysis of the Rabi-type oscillations between two spin states of an atom pair allows a direct determination of the coupling parameters in the interaction hamiltonian. We deduce differences in scattering lengths from our data that can directly be compared to theoretical predictions in order to test interatomic potentials. Our measurements agree with the predictions within 20\%. The knowledge of these coupling parameters allows one to determine the nature of the magnetic ground state. Our data imply a ferromagnetic ground state for \Rb\ in the $f=1$ manifold, in agreement with earlier experiments performed without the optical lattice. For \Rb\ in the $f=2$ manifold the data points towards an antiferromagnetic ground state, however our error bars do not exclude a possible cyclic phase.
\end{abstract}


\maketitle
\tableofcontents

\section{Introduction}
The experimental realization of Bose-Einstein condensation (BEC) in optical traps \cite{Stenger98,Barrett01,Schmaljohann04,Chang04,Kuwamoto04,Higbie05} has opened the possibility to investigate novel effects not accessible in magnetic potentials, where only specific atomic spin states can be trapped. 
In an optical trap the spin-degree of freedom is liberated, allowing to study intriguing phenomena in multi-component gases such as spin dynamics \cite{Schmaljohann04,Chang04,Kuwamoto04}, spin waves \cite{McGuirk02,Gu04}, spin-mixing \cite{Law98,Pu99} and spin-segregation \cite{Lewandowski02}. 
The mechanism behind these phenomena is a coherent interaction between two atoms which changes the individual spins of the colliding particles while preserving the total magnetization \cite{Ho98,Ohmi98,Ciobanu00,Koashi00,Klausen01}.

In the low-energy limit, interatomic interactions are characterized by the corresponding interaction strength, usually parametrized by the $s-$wave scattering length $a$ \cite{Dalibard1999a,Heinzen1999a}. 
In general, this scattering length depends on the spin states in the incoming and outcoming scattering channels.
For two colliding spin-$f$ particles, the interaction strength is characterized by $2f$ independent scattering lengths $a_F$, labelled by the total coupled spin $F$ of the two particles. For bosons (fermions) the total spin $F$ takes even (odd) values between 0 and $2f$. This spin-dependence of the interaction potential is reflected in a many-particle system by a spin-dependent interaction energy. The spin configuration which minimizes this energy at zero magnetic field is energetically favoured and is called the magnetic ground state of the system.

For interacting spin-1 gases this magnetic ground state can be either ferromagnetic or antiferromagnetic in nature, i.e.~spin dependent interactions favour the relative orientation of two atomic spins to be either parallel or antiparallel \cite{Ho98,Ohmi98}. In the spin-2 case an additional cyclic phase can also arise \cite{Ciobanu00,Koashi00,Ueda02}. Here spin-spin correlations are revealed for three particles, whose spin orientations tend to form an equilateral triangle in a plane. These systems are promising in order to study quantum magnetism phenomena \cite{Ho98,Ohmi98,Ciobanu00,Klausen01}, and to create strongly correlated magnetic quantum states in optical traps \cite{Koashi00,Ho00,Pu00,Duan00,Mustecaplioglu02} and optical lattices \cite{Demler02,Yip03,Hou03,Svidzinsky03,Imambekov03,Paredes03,Snoek04,Jin04,Tsuchiya04,GarciaRipoll04}.

Many experiments on multi-component quantum gases involve \Rb\cite{Schmaljohann04,Chang04,Kuwamoto04,Chang05}, the "workhorse" of present day BEC experiments.
Several experiments have investigated the dynamics of a spin-1 \Rb\ Bose-Einstein condensate (BEC) in order to identify the magnetic ground state \cite{Schmaljohann04,Chang04}. There, a slow spin dynamics has been observed on a second timescale leading to a final state close to the predicted ferromagnetic ground state at zero external magnetic field. Recently, an experiment observing coherent spin dynamics in a BEC \cite{Chang05,Kronjaeger05} has shown the spin-1 magnetic ground state of \Rb\ to be ferromagnetic.
For \Rb\ spin-2 atoms the magnetic ground state has been predicted to be antiferromagnetic, but close to the phase boundary of the cyclic phase. Recent experiments \cite{Schmaljohann04,Kuwamoto04}, again observing the spin dynamics for various population of the Zeeman sublevels, seem to agree with this prediction. 
One possible difficulty, however, arises from the fact that even small magnetic fields can perturb the system and pin it to some configuration which is not the ground state at zero external field. Another difficulty in precisely determining the ground state stems from the fact that the differences of the scattering lengths driving the spin dynamics is very small (typically only a few percent of the bare scattering lengths), so that the time scale on which the system relaxes to its ground state is very long.
More recently, a novel method has been proposed theoretically to identify the magnetic ground state of \Rb\ spin-2 atoms \cite{Saito05} circumventing many of the problems described above by observing the time evolution of an initial state with specific population and phases of the Zeeman sublevels.

In this work, we choose another approach and present a method based on the coherence of the spin evolution to directly deduce the absolute value and sign of the various spin-dependent interaction terms in the hamiltonian. From the measured oscillation frequencies we are able to infer the coefficients in the interaction hamiltonian and value of the differences in scattering lengths. The knowledge of these allows in turn to determine the magnetic ground state. There are several advantages of this method. First, our system is intrinsically free from mean-field shifts. Second, the optical lattice leads to a significant enhancement of the coupling strength determining the time scale of the spin-changing dynamics. Therefore, even for the spin-2 case, where strong losses are expected, coherent spin-changing collisions can be observed.
This allows for a direct comparison with calculations based on a recent prediction for \Rb\ ground state $s$-wave scattering lengths. In this sense our measurement constitutes a stringent test of theoretical interatomic interaction potentials.

This paper is structured as follows. Section \ref{sec:Theory} briefly recalls the fundamental mechanism of spin changing collisions and introduces the collisionally induced Rabi-oscillations between two-particle Zeeman states in an optical lattice. We give a detailed description of our experimental sequence in Section \ref{sec:Experiment}. In Section \ref{sec:data} we present the main experimental results and deduce the relevant parameters characterizing the spin-dependent interaction. In Section \ref{sec:DetMagGndState} we describe how the precise determination of the collisional coupling constants can identify the magnetic ground states of spin-1 and spin-2 \Rb\ ground state atoms.

\section{Theory of coherent collisional spin dynamics}\label{sec:Theory}
\subsection{Spin changing collisions in optical lattices}\label{sec:CollisionInOptLat}
Let us consider two colliding spin-$f$ \Rb\ atoms initially in the single-particle Zeeman states with magnetic quantum numbers $m_1$ and $m_2$. The particles can undergo a spin changing collision and be tranferred into a final state characterized by the quantum numbers $m_3$ and $m_4$.
In binary collisions between alkali atoms the total spin projection on the quantization axis is conserved \cite{Ho98,Ohmi98,Law98,Pu99}, which implies $m_1+m_2 = m_3+m_4$. This limits the number of accessible final states to those that have the same total magnetization as the initial one.
The interaction can be described by a potential of the form \cite{Ho98}
\begin{equation}\label{eq:intpot}
	V \left( \vec{r}_1 - \vec{r}_2 \right) = V_s \, \delta \left( \vec{r}_1 - \vec{r}_2 \right) = \sum_{F=0}^{2f} {g_F}\,{\cal P}_F \; \delta \left( \vec{r}_1 - \vec{r}_2 \right) ,
\end{equation}
where $g_F = (4 \pi \hbar^2 /M) \, a_F$, $M$ is the mass of one atom, $\cal{P}_F$ is the projection operator onto the subspace of total spin $F$, $a_F$ is the $s$-wave scattering length for two atoms colliding in such a channel with total spin $F$, and $\hbar=h/2\pi$, with $h$ Planck's constant. For bosons, the total spin $F$ is restricted to even values only \cite{Ho98,Ueda02}. 

In a deep three-dimensional periodic potential (see Section \ref{sec:Experiment}), the situation can be adjusted such that at many ($\sim10^5$) potential minima (lattice sites) there are two atoms trapped in the vibrational ground state of the potential. Throughout the following discussion we assume a sufficiently deep potential such that the atoms can neither leave the trap by tunneling nor be excited via interactions into a higher vibrational state. Consequently, the atoms remain in the lowest vibrational state at all times. Because the spatial dynamics is frozen, the ensuing dynamics caused by spin-dependent interactions only affects the spin component of the two-particle wavefunction. To describe this collisional dynamics, we write two-particle atomic wavefunctions in the form 

\begin{equation}\label{eq:productform}
\ket{\psi} \equiv \ket{\phi_0}_1\ket{\phi_0}_2 \otimes \mathcal{S}\ket{f,m;f,m'}.
\end{equation}
Here $1,2$ labels the two atoms in the pair, $\mathcal{S}$ is the symmetrization operator, \ket{\phi_0} is the spatial wavefunction corresponding to the vibrational ground state in each well. Also \ket{f,m;f,m'} denotes the (unsymmetrized) two-particle spin wavefunction. In the following we describe the atom pair by its spin wavefunction only and imply a constant spatial wave function $\ket{\phi_0}$ that is not explicitly denoted. With this notation, the interaction potential (\ref{eq:intpot}) connecting an initial two-particle state \ket{f,m_1;f,m_2} with a final two-particle state \ket{f,m_3;f,m_4} can be written as 
\begin{equation}
		\bra{f,m_3;f,m_4} V \ket{f,m_1; f,m_2}  =  \tilde{U} \times \Delta a_{m_1,m_2}^{m_3,m_4},
\end{equation}
where the two quantities on the right hand side are given by
 \begin{eqnarray}
	\Delta a_{m_1,m_2}^{m_3,m_4} & =  \sum_{F=0}^{2f}\, \sum_{m=-F}^F \langle f, m_3; f, m_4 \ket{F\, m}\, a_F\\
	\tilde{U} & =  \frac{4 \pi \hbar^2}{M} \times \int\,d^3 r \; \vert \phi_0 \vert^4.\label{eq:intDeltaa}
\end{eqnarray}
Here, $\Delta a_{m_1,m_2}^{m_3,m_4}$ is given by a difference of the scattering lengths $a_F$ weighted by the relevant Clebsch-Gordan coefficients $\langle f, m_3; f, m_4 \ket{F\, m}$ connecting the initial and final spin states through an intermediate, coupled spin state \ket{F\, m}. The pre\-factor $\tilde{U}$ contains all constants and the spatial wavefunction overlap. It can be calculated using
\begin{equation}
	\tilde{U} = \tilde{U}_{\mathrm{HO}} + \tilde{U}_{\mathrm{anharm}} + \tilde{U}_{\mathrm{int}},
\end{equation}
$\tilde{U}_\mathrm{HO}$ is calculated by approximating the lattice potential by a harmonic well near each minimum, and by using the lowest harmonic oscillator eigenfunction as $\phi_0$ \cite{Zwerger03}. In addition, we include two corrections. First, using Wannier functions from a band structure calculation for the wave function $\phi_0$ instead of a gaussian yields a negative correction term $\tilde{U}_\mathrm{anharm}$ which is on the order of $5\%$. 
Second, at short distances the spatial wave function of the colliding particles is expected to deviate from the product form in (\ref{eq:productform}). A preliminary theoretical investigation \cite{MentinkPrivate05} indicates that this deviation may introduce a negative correction $\tilde{U}_{\mathrm{int}}$.
We have experimentally performed a consistency check for the prefactor $\tilde{U}$ by monitoring the collapse and revival of the interference pattern as performed in \cite{Greiner02b}. We measure the interaction matrix element $U$ for the spin state $\ket{f=1,m=-1}$. Assuming a known scattering length for this spin state we infer an approximate value of $\tilde{U}$ consistent within 10\% with calculated values including only the correction due to the anharmonicity $\tilde{U}_{\mathrm{anharm}}$.
Hence, the actual values we use in the following include the first correction term $\tilde{U}_{\mathrm{anharm}}$, but not the second $\tilde{U}_{\mathrm{int}}$. The values are $\tilde{U} = 2 \pi \times (30.4 \pm 1.1)\,\mbox{Hz}/a_\mathrm{B}$ for a lattice depth of $40\,E_r$, and $\tilde{U}=2 \pi \times (33.0 \pm 1.4)\,\mbox{Hz}/a_\mathrm{B}$ for $45\,E_r$. Here $a_\mathrm{B}$ is the Bohr radius, $E_r$ is the single photon recoil energy $E_r = h^2/2M \lambda^2$, with $\lambda$ the wavelength of the lattice laser. The error is given by our uncertainty in optical potential depth, for which we assumed an upper bound of 10\%, but a possible additional systematic error of $\tilde{U}$ due to the correction $\tilde{U}_{\mathrm{int}}$ is not explicitely given.
 Finally it should be noted that the collapse and revival experiment has been performed at a magnetic field of $B=150$\,G in the magnetic trap, whereas the dynamics driven by the spin-dependent interaction occurs at low magnetic fields around or below 1\,G.

\subsection{Two level system}
\begin{figure}[htbp]
	\begin{center}
		\includegraphics[scale=0.45]{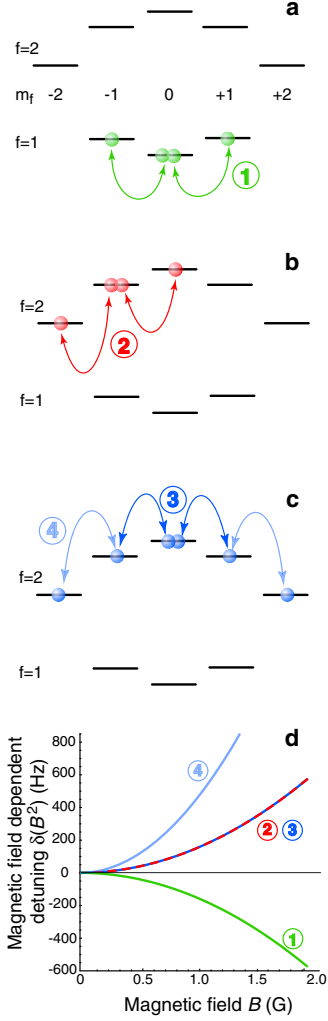}
	\end{center}
	\caption{(a-c) Shift of energy levels due to second order Zeeman effect in the \Rb\ ground state with spin-1 and spin-2. The various processes showing the spin changing dynamics between two-particle Zeeman states are indicated in different colors. (d) Corresponding magnetic field dependent detuning $\delta_B$ for selected combinations of two-particle states versus magnetic field.}
	\label{fig:ZeemanShift2ndOrder}
\end{figure}

In the experiments described below (see cases a-c in Figure \ref{fig:ZeemanShift2ndOrder}), we prepare all atom pairs in the lattice in a specific spin state \ket{\psi_i}. This initial state is in general coupled to one or several final states \ket{\psi_f} by the spin-changing interaction.
In the simplest cases, where the atom pair is initially prepared in \ket{1,0;1,0} (Figure \ref{fig:ZeemanShift2ndOrder}a) or in \ket{2,-1;2,-1} (Figure \ref{fig:ZeemanShift2ndOrder}b), the system can access only one final state, \ket{1,+1;1,-1} or \ket{2,0;2,-2}, respectively, while conserving total magnetization.

As for any two-level system, the probability to find the atom pair in the final state is given by the Rabi formula 
\begin{equation}\label{Eq:RabiFormula}
	P_f=\frac{\Omega^2}{2 \Omega_{if}^{\prime 2}}\left[ 1-\cos(\Omega_{if}^\prime \, t) \right].
\end{equation}
The effective Rabi frequency $\Omega_{if}^\prime$ is given by 
\begin{equation}\label{eq:RabiFreq}
	\Omega_{if}^\prime = \sqrt{\Omega_{if}^2 + \delta_{if}^2},
\end{equation}
where the coupling strength $\Omega_{if}$ is determined by the matrix element $\bra{\psi_f}\hat{V_S}\ket{\psi_i}$ of the spin-changing interaction, and where the detuning $\delta_{if}$ can be decomposed into two parts:
\begin{equation}\label{eq:Detuning}
	\delta_{if} = \delta_{\mathrm{int}} + \delta_{B}.
\end{equation}
The first detuning $\delta_{\mathrm{int}}$ arises from the difference of interaction energies in the initial and final states $\bra{\psi_{i}}\hat{V_S}\ket{\psi_i} - \bra{\psi_{f}}\hat{V_S}\ket{\psi_f}$. The second detuning $\delta_{B}$ is due to the energy shift of initial and final states in a non-zero magnetic field. 
Since the total magnetization is conserved, the system is not sensitive to first order Zeeman shifts, because the initial and the final states experience the same first order Zeeman effect \cite{Stamper-Kurn00}. Consequently the detuning is to leading order determined by the quadratic Zeeman shift. Therefore the term $\delta_B$ in Equation (\ref{eq:Detuning}) is given by the second-order Zeeman-effect $\delta_B \propto \frac{(\mu_B B)^2}{2 \hbar \omega_{\mathrm{hfs}}}$, where $\omega_{\mathrm{hfs}}$ denotes the hyperfine splitting. The corresponding detuning $\delta_\mathrm{B}$ is shown versus magnetic field in Fig.~\ref{fig:ZeemanShift2ndOrder}d. 

As indicated in the previous section, the two parameters of the Rabi model can be written in the form $\Delta a \times \tilde{U}$, where the value of $\Delta a$, different for $\Omega_{if}$ and $\delta_{if}$, can be calculated for each specific case (see Table \ref{tab:RabiSummary}). 
It should be noted that for \Rb\ the bare scattering lengths $a_F$ are almost equal; recently calculated values \cite{vanKempen02} are  $a_0=(101.78 \pm 0.2)\,a_\mathrm{B}$ and $a_2=(100.40 \pm 0.1)\,a_\mathrm{B}$ for spin-1. Predicted values for spin-2 \cite{MentinkPrivate05} are $a_0 = (87.93 \pm 0.2)\,a_\mathrm{B}$, $a_2 = (91.28 \pm 0.2)\,a_\mathrm{B}$ and $a_4 = (98.98 \pm 0.2)\,a_\mathrm{B}$, where the Bohr-radius $a_\mathrm{B} = 5.29\times10^{-11}\,$m. Thus the differences in scattering lengths describing the spin-dependent coupling give values of $\Delta a$ on the order of a few percent of the bare scattering lengths $a_F$.

Since the dependence of the magnetic field dependent detuning with external field is known, a measurement of the effective Rabi frequency $\Omega_{if}^\prime$ for different magnetic fields allows to extract the absolute value of the coupling strength $\Omega_{if}$ and both the absolute value of the constant detuning $\delta_{\mathrm{int}}$ and its sign.
The form of Equation (\ref{eq:RabiFreq}) implies that if $\delta_{\mathrm{int}}$ and $\delta_B$ have the same sign,
even at zero external magnetic field Rabi oscillations do not show unity contrast, because the constant detuning $\delta_{\mathrm{int}}$ is present at all times. This can be overcome e.g.~applying a microwave field which can shift the two two-level states with respect to each other in energy. The effect of this AC Zeeman effect on spin dynamics in an optical lattice has been described in detail in \cite{Gerbier05b}.

\subsection{Three level system} \label{sec:ThreeLevel}
The situations investigated in the previous section were describable as pure two-level systems.
Another possibility (case (c) in Figure \ref{fig:ZeemanShift2ndOrder}) is that two final states $\ket{\psi_{f_1}}$ and $\ket{\psi_{f_2}}$ fulfill the condition of a conserved total magnetization.
The dynamics can then be modelled by a coupled three-level system, as has been calculated in detail in \cite{Huang05}.
Assuming that the direct coupling between \ket{\psi_i} and \ket{\psi_{f_2}} is negligible, the corresponding interaction hamiltonian reads
\begin{equation}\label{eq:threelevel}
	V = \frac{\hbar}{2} \left ( \begin{array}{ccc} 0       & \Omega_1     & 0 \\
																								\Omega_1 & 2\, \delta_1 & \Omega_2 \\
																								0        & \Omega_2     & 2\, \delta_2
																								 \end{array} \right),
\end{equation}
where $\Omega_i$ and $\delta_i$ describe the coupling strength and total detuning, respectively, and $i=1(2)$ labels the process $\ket{\psi_i} \leftrightarrow \ket{\psi_{f_1}}$ ($\ket{\psi_{f_1}} \leftrightarrow \ket{\psi_{f_2}}$). Similar to the treatment in \cite{Fewell97} we find the eigenfrequencies $\omega$ as solutions to the secular equation
\begin{equation}
	\omega^3 - \omega^2 (\delta_1 + \delta_2) + \omega \left (\delta_1 \delta_2 - \frac{\Omega_1^2}{4} - \frac{\Omega_2^2}{4} + \frac{\Omega_1^2}{4}\; \delta_2 \right ) = 0.
\end{equation}
The eigenfrequencies then read
\begin{eqnarray}
	\omega^0 &=& \frac{1}{3} \left (\delta_1 + \delta_2 + \tilde{\Omega}\, \cos\frac{\zeta}{3} \right ) \nonumber \\
	\omega^+ &=& \frac{1}{3} \left (\delta_1 + \delta_2 + \tilde{\Omega}\, \cos\frac{2 \pi - \zeta}{3} \right ) \nonumber \\
	\omega^- &=& \frac{1}{3} \left (\delta_1 + \delta_2 + \tilde{\Omega}\, \cos\frac{2 \pi + \zeta}{3} \right ),\label{eq:ThreeLevelModel}
\end{eqnarray}
where 
\begin{eqnarray}
	\tilde{\Omega} &=& \sqrt{3 (\Omega_1^2 + \Omega_2^2) + 4 (\delta_1^2 - \delta_1 \delta_2 + \delta_2^2)} \nonumber \\
	\zeta          &=& 2 \pi - \arccos \left[ \frac{1}{\tilde{\Omega}^3}\;[9 (\Omega_1^2 + \Omega_2^2) \right. \nonumber \\
	&&\left.- 4 (2 \delta_1 - \delta_2)(2 \delta_2 - \delta_1)](\delta_1 + \delta_2) - 27 \Omega_1^2 \delta_2 \right]. \label{eq:ThreeLevelFreq}
\end{eqnarray}
In the case where \ket{\psi_{f_2}} is separated by a very large energy difference from the other two states, $\delta_2 \gg \delta_1$, the system effectively reduces to a two-level system oscillating between $\ket{\psi_i}$ and $\ket{\psi_{f_1}}$. The relevant oscillation frequency then becomes $\lim_{\delta_2 \to \infty} (\omega^0 - \omega^-) = \sqrt{\Omega_1^2 + \delta_1^2}$. In this limit, a description as a two-level system $\ket{\psi_i}\leftrightarrow \ket{\psi_{f_1}}$ is valid.

Comparing to the situation depicted in Figure \ref{fig:ZeemanShift2ndOrder}c this means that for low magnetic fields ($B\lesssim 0.6\,$G) a model including all three states has to be used in order to describe the dynamics. For increasing magnetic fields, the second order Zeeman shift detunes the final state $\ket{\psi_{f_2}}=\ket{2,+2;2,-2}$ faster than the other final state $\ket{\psi_{f_1}}=\ket{2,+1;2,-1}$ (see Figure \ref{fig:ZeemanShift2ndOrder}d), such that for large enough magnetic fields ($B\gtrsim 0.6\,$G) a two-level model only including two states \ket{2,0;2,0} and \ket{2,+1;2,-1} is sufficient.

\section{Experimental sequence}\label{sec:Experiment}
\begin{figure}[htbp]
	\begin{center}
		\includegraphics[scale=0.6]{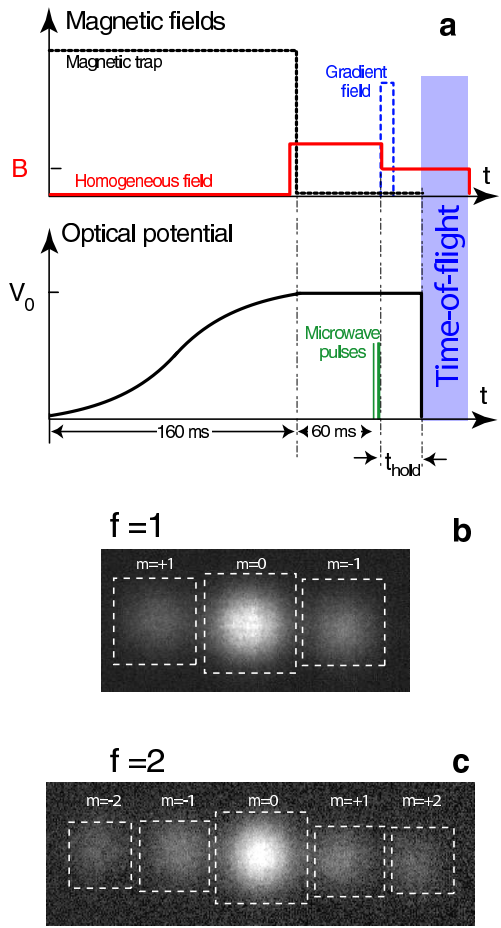}
	\end{center}
	\caption{(a) Sketch of the experimental sequence. (b) Typical camera image for atoms in $f=1$ and (c) $f=2$. The dashed boxes illustrate areas in which the atom number for each magnetic sublevel is counted.}
	\label{fig:ExperimentalSequence}
\end{figure}

In order to realize the different situations of spin-dynamics in an optical lattice as depicted in Figure \ref{fig:ZeemanShift2ndOrder}(a-c), we start our experimental sequence by creating a \Rb\ Bose-Einstein Condensate (BEC) containing around $2\times 10^5$ atoms in the \ket{f=1; m_f=-1} hyperfine state in an almost isotropic magnetic trap (trap frequency $\omega_M \approx 2 \pi \times 15$\,Hz). We then load the sample into a combined magnetic and optical trap by slowly increasing the power of three mutually orthogonal standing waves to a final trap depth of either $40\,E_r$ or $45\,E_r$, driving the system into the Mott-insulator regime \footnote{The lattice depths in our experiment are calibrated by modulating the depth slightly (20\,\%) at a frequency $\nu_\mathrm{mod}$ and by monitoring the resonance excitation to the second band \cite{Jauregui01}. The lattice depth is deduced by comparing $\nu_\mathrm{mod}$ to a band-structure calculation \cite{GreinerPhD03}.}. Here $E_r$ is the single photon recoil energy $E_r = h^2/2 M \lambda^2$, with $\lambda=840\,$nm the lattice laser wavelength.
In the later discussion of systematic errors we assume an uncertainty in the potential depth of $\sim10\%$, which should be understood as an upper bound. The particular form of the lattice-intensity ramp has been taken from \cite{Jaksch02}, minimizing excitations in the system \cite{Gericke06}. It has a total duration of $160$\,ms (see Fig.~\ref{fig:ExperimentalSequence}).

In order to observe spin oscillations, the magnetic trap is switched off, leaving the sample in a pure optical trap. The spin-polarization of the atoms is preserved by a small homogeneous offset-field of $\sim1.2\,$G which is switched on 5\,ms before switching off the magnetic trap (see Fig.\ref{fig:ExperimentalSequence}).
After a hold time of 60\,ms during which the bias field of the magnetic trap decays to below 1\,mG, the atoms are prepared in their initial state for spin dynamics by application of microwave pulses with a frequency around $6.8$\,GHz and a pulse duration between $6\,\mu$s and $150\,\mu$s, depending on the specific transition. 

Directly after the last pulse, the homogeneous offset field is ramped to its final value. The time constant of the magnetic field change is on the order of a few hundred $\mu$s; this magnetic field has been calibrated by microwave spectroscopy and is known to a value better than $\sim10\,$mG, which in the following we consider as an upper bound of the systematical error in the magnetic field.
After a variable hold time $t$ the optical trapping potential is switched off and the atoms can expand during 7\,ms of time-of-flight (TOF), while the homogeneous magnetic field is still present. During the first 3\,ms of TOF a gradient field is switched on in order to spatially separate the different magnetic sublevels. The population $N_m$ of each magnetic sublevel $m$ is recorded after TOF with standard absorption imaging by counting the fractions $N_m$ in a small region around the corresponding atom cloud  (see dashed boxes in Figures \ref{fig:ExperimentalSequence}b,c). The total atom number $N_\mathrm{tot}$ is counted in a large region including all magnetic sublevels. However, the different atom clouds corresponding to different magnetic sublevels are not well enough separated. In order to avoid double counting of atoms, we chose counting regions that do not overlap, but slightly underestimate the population $N_m$ of each cloud. Moreover, an additional background signal on the order of a few percent is counted in the global region used to find the total atom number $N_\mathrm{tot}$.  The populations in different Zeeman states as given later therefore sum up to a value slightly ($<5\%$) smaller than 1.

Due to the gaussian shape of the lattice laser beams, the atoms in the optical lattice experience an additional smooth confinement and therefore the system favours the formation of Mott shells with distinct, integer number of atoms per site \cite{Jaksch98,Batrouni02,Gerbier05}. 
For our trap parameters we calculate from a mean field model that a central core of atom pairs contains approximately half of the atoms \cite{Gerbier05a}. This core is surrounded by a shell with isolated atoms, whereas the number of sites with more than two atoms is negligible. 
Due to this shell structure, not all atoms contribute to the spin oscillations.
Consequently the measured amplitude of the spin oscillation -- normalized to the signal from \emph{all} atoms -- is decreased compared to the value predicted by the Rabi model. The measured relative amplitude is expected to be
\begin{equation}
	\frac{N_{+1} + N_{-1}}{N_\mathrm{tot}} = n \, P_f,
\end{equation}
where $n$ is the fraction of atoms in sites with double filling. 

\section{Experimental results}\label{sec:data}
We have experimentally investigated the spin dynamics in the three situations depicted in Figure \ref{fig:ZeemanShift2ndOrder}(a-c), where the system is initialized in different internal atomic states. The initial states for those cases are (i) \ket{f=1,m_f=0}, (ii) \ket{f=2,m_f=-1}, and (iii) \ket{f=2,m_f=0}. In most cases, the assumption that for an initial state only one final state is accessible after a spin changing collision is valid. In the lower $f=1$ hyperfine manifold, this is naturally fulfilled, because only one combination of two-particle states conserves total magnetization. For the upper $f=2$ hyperfine state, there are in general more initial and final states accessible. For case (ii) the description as a two-level system is valid at all parameters used in our experiments. For case (iii), as already discussed in Section \ref{sec:ThreeLevel}, a sufficiently large external magnetic field can be used to tune the system via the quadratic Zeeman-shift into a regime where the approximation by a two-level system is sufficient. For small magnetic fields, however, a description as a three-level system must be considered.

Another difference between the $f=1$ and $f=2$ cases is the role of losses. Due to our preparation in the optical lattice, three-body recombination processes are suppressed, because most atoms are either in singly or doubly occupied lattice sites. In the $f=1$ state, two-body loss rates are very low, so that atom losses are negligible. This is not the case in $f=2$ where hyperfine changing collisions have been seen to be significant \cite{Burt97,Myatt97,Schmaljohann04} and to lead to a dephasing of the spin oscillations \cite{Widera05}. We still observe several coherent spin-oscillations in the upper hyperfine state, because of the relatively large spin-dependent coupling strengths in $f=2$. 

In each case, the frequency of the coherent spin oscillations can be changed through the magnetic field dependent detuning $\delta_B$. We record the population in each Zeeman sublevel as a function of time for the three initial states introduced above for different values of the external magnetic field and extract the corresponding oscillation frequency.
For each process we fit the data to the expected behavior of the oscillation frequency (\ref{eq:RabiFreq}). Since the spin-dependent dynamics is well described by one (two) differences in scattering lengths for the $f=1$ ($f=2$) case, we fit the oscillation frequency vs magnetic field behaviour with only one (two) free parameters corresponding to the scattering length differences. In order to do this, we make use of the relations given in Table \ref{tab:RabiSummary}, re-parametrizing the coupling strengths $\Omega_{if}$ in terms of the detunings $\delta_{\mathrm{int}}$; hence, we use the fitting parameter $\delta_1$ for $f=1$, being the interaction detuning $\delta_{\mathrm{int}}$ for the process $\ket{1,0;1,0} \leftrightarrow \ket{1,+1;1,-1}$. For $f=2$ we use the two interaction detunings $\delta_{2,1}$ for the process $\ket{2,0;2,0}\leftrightarrow \ket{2,+1;2,-1} \leftrightarrow \ket{2,+2;2,-2}$ and $\delta_{2,2}$ for $\ket{2,-1;2,-1} \leftrightarrow \ket{2,0;2,-2}$ as fitting parameters.  
We do this instead of directly fitting the scattering length differences, because the factor $\tilde{U}$ linking the resulting detunings $\delta_{\mathrm{int}}$ to the scattering length differences $\Delta a$ is not error-free and might be systematically shifted. Finally, from the extracted detunings $\delta_{\mathrm{int}}$ and the calculated values for $\tilde{U}$ (see Section \ref{sec:CollisionInOptLat}), we give values for the scattering length differences calculated from the extracted detunings $\delta_{\mathrm{int}}$.
 
\subsection{Dynamics in the $f=1$ hyperfine state}
\begin{figure}[htbp]
	\begin{center}
		\includegraphics[scale=0.75]{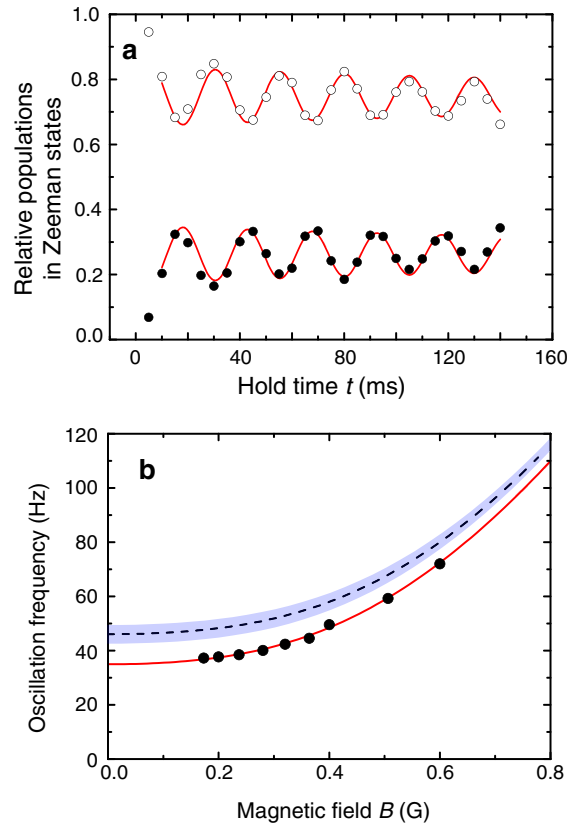}
	\end{center}
	\caption{(a) Spin population oscillations in $f=1$ between \ket{1,0;1,0} ($\medcirc$) and \ket{1,+1;1,-1} ($\medbullet$) at a magnetic field of 0.28\,G and a lattice depth of $45\,E_r$. A fit to a damped sine (solid lines) yields an oscillation frequency of 40.2(3)\,Hz and a damping rate of $\gamma = 3(2)\,$Hz. (b) Measured oscillation frequency of the $\ket{1,0;1,0} \leftrightarrow \ket{1,+1;1,-1}$ collision process versus external magnetic field $B$. The solid line is a fit to the expected behaviour according to the effective Rabi-frequency $\Omega_{if}^\prime$. The dashed line is the calculated curve based on the predicted scattering lengths \cite{vanKempen02}. The shaded region reflects our uncertainty in magnetic field of $\pm 10\,$mG and the uncertainty in potential depth (upper bound of 10\%).}
	\label{fig:f=1Oscillation_combined}
\end{figure}

The collisional coupling between two $f=1$ atoms is an almost ideal example of the Rabi model introduced above. 
The only possible process conserving total magnetization is $\ket{1,0;1,0} \leftrightarrow \ket{1,+1;1,-1}$. 
In order to observe spin dynamics, the atomic sample is prepared by a first microwave pulse in the \ket{f=2; m_f=0}-state and immediately a second pulse is applied, transferring the atoms into \ket{f=1,m_f=0}. A typical oscillation for a lattice depth of $45\,E_r$ at a magnetic field of $B=0.28\,$G is shown in Fig.~\ref{fig:f=1Oscillation_combined}a. 
Spin oscillations have been observed for various magnetic fields between approximately 170\,mG and 600\,mG. For larger magnetic fields, the amplitude is strongly suppressed, because the magnetic field dependent detuning $\delta_B$ becomes large compared to the coupling strength. In this limit, spin oscillations become difficult to resolve.
The extracted interaction detuning (see Section \ref{sec:appendixRabi}) is $\delta_{\mathrm{int}}(f=1) \equiv \delta_{1} = 2 \pi \times (-11.8^{\pm 0.04}_{\pm 0.2})\,$Hz, where the upper error gives the statistical uncertainty, and the lower error the systematic uncertainty in lattice depth. 

\subsection{Dynamics in the $f=2$ hyperfine state}
For the spin-2 case, coherent spin oscillations driven by interatomic collisions in an optical lattice have been presented in \cite{Widera05}. Here we re-analyse the data, including a description by a coupled three-level model.
\paragraph{Two-level system.}
For the case where the dynamics starts from \ket{2,-1;2,-1}, the only possible interconversion process that preserves total magnetization is $\ket{2,-1;2,-1} \leftrightarrow \ket{2,0;2,-2}$. Thus this situation can also be described as a two-level system.  

\paragraph{Three-level system.}
For the case where the system is initialized in \ket{2,0;2,0}, there are, however, two final states \ket{2,+1;2,-1} and \ket{2,+2;2,-2} accessible. The direct process $\ket{2,0;2,0} \leftrightarrow \ket{2,+2;2,-2}$ is strongly suppressed, but this final state can be populated through a two-step process $\ket{2,0;2,0} \leftrightarrow \ket{2,+1;2,-1} \leftrightarrow \ket{2,+2;2,-2}$. For sufficiently large magnetic field, the state \ket{2,+2;2,-2} is far detuned so that the situation is effectively described by a two-level system \cite{Widera05}. 

\begin{figure}[htbp]
	\begin{center}
		\includegraphics[scale=0.5]{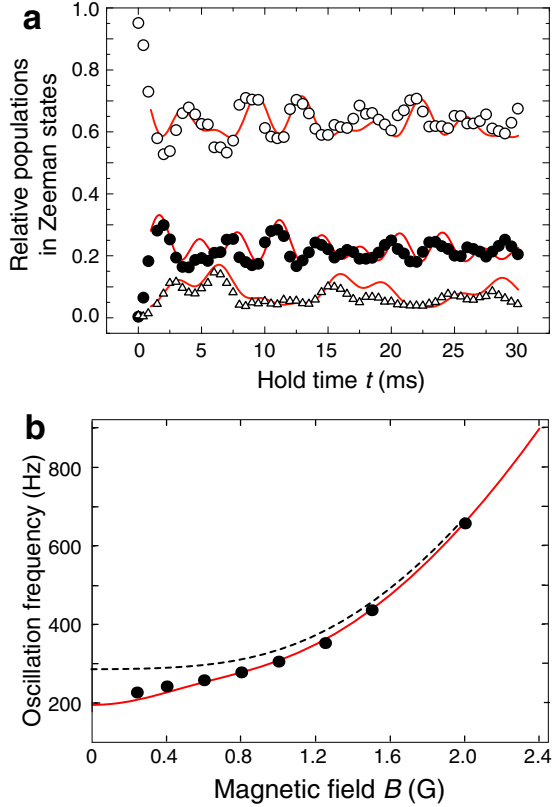}
	\end{center}
	\caption{(a) Measured population dynamics for the coupled three-level system $\ket{2,0;2,0}\leftrightarrow \ket{2,+1;2,-1} \leftrightarrow \ket{2,+2;2,-2}$ at a magnetic field of $B=0.24$\,G and a lattice depth of $40\,E_r$. One clearly sees the deviation from the pure sinusoidal two-level oscillations and the rather large population of \ket{2,+2;2,-2}. The solid lines are calculations based on a coupled three-level system (\ref{eq:threelevel}) with parameters corresponding to the experimental realization. (b) Oscillation frequency vs magnetic field for spin dynamics starting from \ket{2,0;2,0}. The red line is a fit to the analytical solution of the three-level model (\ref{eq:ThreeLevelModel}), the dashed line shows the result of a two-level model as presented in \cite{Widera05}.}
	\label{fig:ThreeLevels}
\end{figure}

For lower magnetic fields, a three-level description as in Section \ref{sec:ThreeLevel} must be used in order to describe the population oscillation. In Figure \ref{fig:ThreeLevels}a we show a population oscillation for a lattice depth of $40\,E_r$ and a magnetic field of $B=0.24\,$G which clearly differs from the simple sinusoidal case. Here, the solid lines are a calculation from a three-level model which shows good agreement with the measured data.
For this calculation we use the fitted values of the two detunings $\delta_{2,1}$ and $\delta_{2,2}$ (see below) together with the form of the Rabi parameters given in Table \ref{tab:RabiSummary}. With this we calculate the values of the three-level model (\ref{eq:threelevel}) $\Omega_{1} = 2\pi \times 265,$Hz, $\delta_{1}=2 \pi \times 33\,$Hz for the coupling of $\ket{2,0;2,0}$ with \ket{2,+1;2,-1}, and $\Omega_{2} = 2 \pi \times 134\,$Hz, $\delta_{2} = 2 \pi \times 100\,$Hz for the coupling between \ket{2,+1;2,-1} and \ket{2,+1,2,-2}. 
The phenomenological damping rates for both processes have been set to $\gamma = 30\,\mbox{s}^{-1}$ as measured in \cite{Widera05} and the decay from \ket{2,+2;2,-2} has been neglected. The amplitude has been set to account for $\sim50\%$ of the atoms contributing to the spin-dynamics, and the offset for each population (see Section \ref{sec:Damping}) has been set in order to match the data.

Although the coupled three-level system defined by (\ref{eq:threelevel}) can in principle be solved, we were not able to fit its solution reliably to the data because of the large number of parameters involved. Instead, we fit the $m_f = 0$ population with the same sinusoidal form as used in the previous cases to approximate the predominant oscillation frequency.  We have checked that a Fourier analysis yields oscillation frequencies in agreement with those found by this fitting procedure. The observed oscillation frequency versus magnetic field is shown in Figure \ref{fig:ThreeLevels}b.
In order to include the three-level character, we fit the data for this process with the analytical solution of the three-level model. The fit is shown as solid line in Figure \ref{fig:ThreeLevels}b, including the dependency of this process on the other process $\ket{2,-1;2,-1} \leftrightarrow \ket{2,0;2,-2}$ via the interaction detuning $\delta_{2,2}$. The theoretical prediction from a two-level system (as presented in \cite{Widera05}) is shown as dashed-dotted line and does not describe the dynamics well for small magnetic fields. A theoretical prediction including the third level, however, does reproduce the correct behaviour (dashed line in Figure \ref{fig:ThreeLevels}b).

\subsection{Determining scattering length differences}\label{sec:ScatLengthDif}
For the analysis it is important to notice that the dynamics in a spin dependent collision of two spin-1 (respectively spin-2) atoms is determined by only one (respectively two) scattering length difference(s). As can be seen from Table \ref{tab:RabiSummary}, the parameters $\Omega_{if}$ and $\delta_\mathrm{int}$ extracted from a fit to the measured change of oscillation frequency versus magnetic field are directly related to those scattering length differences.

For $f=1$ the difference $(a_2-a_0)$ directly results from a fit of the form (\ref{eq:RabiFreq}) to the measured data. We perform this fit with one free parameter $\delta_1$, the interaction detuning for the process. The fit yields a value of $\delta_1 = 2\pi \times (-11.8^{\pm 0.04}_{\pm 0.2})\,$Hz. We stress, that this parameter is still free from systematic effectes due to our uncertainty in optical potential depth, but has a statistical error, as well as a systematic error due to magnetic field uncertainty. The interaction strength $\Omega_{if}$ can be expressed in terms of this parameter as $\Omega_{if} = 2 \sqrt{2} \delta_1$.

For $f=2$ the situation is more complicated. In particular, the process $\ket{2,-1;2,-1} \leftrightarrow \ket{2,0;2,-2}$ depends only on the scattering length difference $(a_4 - a_2)$, whereas the dynamics starting from \ket{2,0;2,0} depends on both differences $(a_4 - a_2)$ and $(a_2 - a_0)$. We reanalyze the data presented in \cite{Widera05}. There, the dynamics was treated in a pure two-level model, and the Rabi-parameters were considered as independent. Here, we first fit the data set for oscillation frequency vs magnetic field of the process $\ket{2,-1;2,-1} \leftrightarrow \ket{2,0;2,-2}$ with only one parameter $\delta_{2,2}$, reflecting the scattering length difference $(a_4 - a_2)$. The resulting best fit parameter is $\delta_{2,2}=2\pi \times (30.2^{\pm 0.1}_{\pm 0.3})\,$Hz.
Then we fit the data set for the dynamics starting from \ket{2,0;2,0} with $\delta_{2,2}$ fixed to this value, and leave the second relevant interaction detuning $\delta_{2,1}$ --- reflecting the scattering length difference $(a_2 - a_0)$ --- as free parameter. For this fit, which is shown in Figure \ref{fig:ThreeLevels}b, we employ the analytical solution of the three level system Equations (\ref{eq:ThreeLevelFreq}). The best fit is achieved for $\delta_{2,1}=2\pi \times (33.4^{\pm 1.4}_{\pm 1.1})\,$Hz.
Due to the small uncertainty in each data point of the first data set compared to the uncertainty of the second data set, this treatment is equivalent to one combined fit of both data sets with two parameters.

In order to extract the scattering length differences, we employ Table \ref{tab:RabiSummary} and the given values for $\tilde{U}$. Those scattering length differences are summarized in Table \ref{tab:Summary}. The values now have an uncertainty due to the optical potential, entering via the constant $\tilde{U}$.
\begin{table}[htbp]
	\begin{center}
		\begin{tabular}{|c|c|c|c|}\hline
					$f$& $\Delta a$ & Measured  & Predicted \\ \hline \hline
					$f=1$ & $(a_2 - a_0)$ & $\left (-1.07 ^{\pm 0.01}_{\pm 0.02 \, \pm 0.06} \right) \, a_\mathrm{B}$ & $-1.38\,a_\mathrm{B}$  \\ \hline
					      & $(a_2 - a_0)$ & $\left (3.51 ^{\pm 0.2}_{\pm 0.18 \, \pm 0.16} \right) \, a_\mathrm{B}$ & $3.35\,a_\mathrm{B}$   \\ \cline{2-4}
						\raisebox{1.5ex}[-1.5ex]{$f=2$} & $(a_4 - a_2)$ & $\left (6.95 ^{\pm 0.02}_{\pm 0.07 \, \pm 0.26} \right) \, a_\mathrm{B}$ & $7.70\,a_\mathrm{B}$   \\ \hline 
		\end{tabular}
	\end{center}
	\caption{Scattering length differences inferred from the spin-oscillations measurement. The errors in the upper row give the statistical uncertainty from the fit, whereas the first and second value in the lower row give the systematic error due to the uncertainty in magnetic field and optical potential depth, respectively. The theoretical predictions are calculations based on values from \cite{vanKempen02,MentinkPrivate05}.}
	\label{tab:Summary}
\end{table}

\subsection{Damping mechanisms}\label{sec:Damping}
An experimentally observed feature of the coherent spin dynamics in both hyperfine states is the damping which is not expected from the models describing the coherent dynamics. For the $f=2$ manifold and high lattice depths the damping rate coincides with the loss rate of atoms due to hyperfine changing collisions \cite{Widera05}.
Although this loss process is absent for $f=1$, a finite damping on the order of $\gamma \approx 3\,\mbox{s}^{-1}$ is observed even in the spin-1 data. One source for this damping could be the inhomogeneity of our system. Due to the gaussian intensity profile of the lattice laser beams, the wavefunction overlap $\int \, d^3r\vert \phi_0 \vert^4 $ is slightly higher in the center than in the outer regions of the system. Therefore the coupling strength $\Omega_{if}$ and the constant detuning $\delta_{\mathrm{int}}$ are position dependent. This leads to a small dephasing in the system, with a rate that we estimate to be on the order of $0.3\,\mbox{s}^{-1}$. Therefore we can exclude its influence on the observed damping.
Another possible explanation for the observed damping at high lattice depths in the spin-1 case are off-resonant Raman transitions introduced by the lattice laser beams. The expected rate of those events has been estimated to be similar to the observed damping rate. However, this effect has not been experimentally investigated. Also, heating to higher vibrational bands and subsequent tunneling to neighboring sites is a possible damping mechanism.

Another feature of the coherent spin dynamics in $f=1$ that has also been observed in $f=2$ is an initial rise of the population in the final two-particle state, which is faster than predicted by the simple Rabi-model.
This fast initial rise at the beginning of the spin dynamics together with the offset of the oscillations can partially be explained by the magnetic field ramp after the last microwave pulse. Here, the sample experiences a changing detuning, and the state from which the spin dynamics start from can be changed. A calculation including a model of the ramped magnetic field yields an offset of the spin population oscillations which is a few 10\% of the experimentally observed offset.

\section{Nature of the magnetic ground state}\label{sec:DetMagGndState}
As explained in the introduction, the relative orientation of two spins in a collisional event changes the actual interaction strength. In order to determine the particular orientation that minimizes the mean field energy at zero magnetic field, i.e.~the magnetic ground state, it is convenient to express the interaction potential in terms of spin-independent and spin dependent parts (see e.~g.~\cite{Ho98,Ueda02}).

For the spin-1 case this yields the potential 
\begin{equation}\label{eq:int_f=1}
	V_S = c_0 + c_2 \vec{f}_1\cdot \vec{f}_2,
\end{equation}
where $c_0 \equiv (4\pi\hbar^2/M) \times (a_0 + 2 a_2)/3$ is the spin-independent part, and $c_2 \equiv (4\pi\hbar^2/M) \times (a_2 - a_0)/3$ describes spin-spin interactions. 
For the antiferromagnetic or polar phase, the mean field energy according to (\ref{eq:intpot}) is minimized by aligning the spins of two interacting atoms anti-parallel, and it emerges for $c_2>0$, i.e.~$a_2 > a_0$ \cite{Ho98}, whereas the mean field energy in the ferromagnetic case is minimized by aligning the spins parallel, implying $c_2<0$. 

\begin{figure}[htbp]
	\begin{center}
		\includegraphics[scale=0.35]{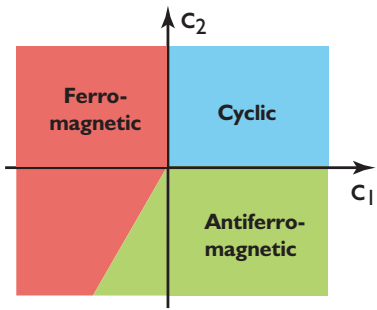}
	\end{center}
	\caption{Phase diagram of the $f=2$ magnetic ground state in \Rb. }
	\label{fig:PhaseDiagramm}
\end{figure}

In a similar way, the interaction potential for the spin-2 case  can be written as \cite{Ueda02}
\begin{equation}\label{eq:int_f=2}
	V_S = c_0 + c_1 \vec{f}_1 \cdot \vec{f}_2 + 5 c_2 {\cal P}_{0}.
\end{equation}
Here, ${\cal P}_{0}$ is a projector onto the singlet subspace, $c_0 \equiv (4\pi\hbar^2/M) \times (4 a_2 + 3 a_4)/7$ describes the spin-independent part, $c_1 \equiv (4\pi\hbar^2/M) \times (a_4 - a_2)/7$ determines the spin-spin interaction and $c_2 \equiv (4\pi\hbar^2/M) \times (7 a_0 - 10 a_2 + 3a_4)/7$ accounts for the interaction between spin-singlet pairs. 
Differently from the spin-1 case, there exist three possible phases for a spin-2 Bose-gas at zero magnetic field. In addition to the ferromagnetic and antiferromagnetic phases the system can be in a cyclic phase. In this phase the spin orientation of three interacting particles tend to form an equilateral triangle in a plane. The phase diagram is now two dimensional, depending on the spin-dependent coefficients $c_1$ and $c_2$. 
A sketch of the phase diagram is given in Figure \ref{fig:PhaseDiagramm}.
The ground state is ferromagnetic for $c_1<0$ and $c_1 - c_2/20<0$, antiferromagnetic for $c_2<0$ and $c_1 - c_2/20>0$, and cyclic for $c_1>0$ and $c_2>0$. 

A summary of the values for the spin-dependent coefficients in $f=1$ and $f=2$ determined from the frequency measurement of the coherent spin dynamics is given in Table \ref{tab:CoefficSummary} together with the theoretically predicted values.
\begin{table}[htbp]
	\begin{center}
		\begin{tabular}{c|c|c|c}
			$f$ & $c_i$  & Measured $\Delta a/a_\mathrm{B}$               & Theoretical $\Delta a/a_\mathrm{B}$ \\ \hline \hline
			$f=1$      &  $c_2/(4\pi\hbar^2/M)$ & $-0.36^{\pm0.01}_{ \pm0.01 \pm 0.02}$ & $-0.46 $ \\ \hline
			           &  $c_1/(4\pi\hbar^2/M)$ &  $+0.99^{\pm 0.01}_{\pm0.01 \pm 0.04}$ & $+1.1 $ \\ \cline{2-4}
			\raisebox{1.5ex}[-1.5ex]{$f=2$} & $c_2/(4\pi\hbar^2/M)$ &  $ -0.53^{\pm 0.2}_{\pm 0.18\pm 0.2}$ & $-0.05$
		\end{tabular}
	\end{center}
	\caption{Summary of the measured coefficients of the spin-dependent interaction (\ref{eq:int_f=1},\ref{eq:int_f=2}). The errors are the statistical errors (upper row) and the systematic errors (lower row, first number due to magnetic field, second due to lattice potential). The calculated values are based on recently predicted scattering lengths \cite{vanKempen02,MentinkPrivate05}.}
	\label{tab:CoefficSummary}
\end{table}
For the spin-1 case the inferred coefficient $c_2$ clearly shows a ferromagnetic ground state at zero magnetic field. 
 For the spin-2 case the situation is not so clear. The coefficient $c_1$ clearly excludes a ferromagnetic ground state, and the coefficient $c_2$ indicates an antiferromagnetic ground state, but the error bars include the phase boundary to the cyclic phase. Moreover, we have observed that even slight changes in, e.g., the fit procedure can change the sign of $c_2$. We conclude that although the data measured data strongly points towards an antiferromagnetic ground state, our analysis does not allow a final determination of the ground state in $f=2$.

\section{Conclusion}
In summary, we have shown that both hyperfine states of \Rb\ show collisionally driven spin oscillations between two-particle states in a deep optical lattice. The oscillations can be described by a Rabi-model for a broad range of parameters. In the case where two final states are accessible, the observed spin dynamics can be explained by a coupled three-level system.

The parameters of the Rabi model are directly related to differences of the scattering lengths in the \Rb\ hyperfine states $f=1$ and $f=2$. The measured values can be compared to calculations based on recent theoretical predictions of the scattering lengths, and show good agreement. We observe, however, that a discrepancy on the order of 10\% persists in all cases (see also \cite{Widera05}), which cannot be explained by either the experimental error bars or the uncertainty in the theoretically predicted scattering lengths.
This discrepancy may have two reasons. First, our approach of using non-interacting wave functions as introduced in Section \ref{sec:CollisionInOptLat} in order to describe the wave function overlap might break down for deep lattices. This problem still requires further theoretical investigation. It should be noted, however, that the conclusions drawn in Section \ref{sec:DetMagGndState} remain unaffected by an overall change of $\tilde{U}$. The reason is that all arguments concerning the sign of the coefficients $c_1$ and $c_2$ hold equally for the measured interaction detunings $\delta_{\mathrm{int}}$ of the different processes, as long as the factor $\tilde{U}$ is independent of the specific spin wave function.

A second origin of the observed discrepancy between the experiment and the theoretical prediction might be that the description of the dynamics by only the spin-changing interaction potential Equation (\ref{eq:intpot}) is not valid. 
For example, dipole-dipole interaction, although weaker than the spin-changing interaction, is also present in principle. In a perfectly isotropic trap, they do not change the interaction energy. In an anisotropic lattice well, however, effects due to dipole-dipole interactions might affect the observed population oscillations \cite{YouPresentation06}. In this case, the conclusions drawn concerning the magnetic ground state of the two hyperfine manifolds might change \cite{Yi04}.
This unresolved problem therefore needs further experimental and theoretical investigation.

\acknowledgements
We would like to thank Servaas Kokkelmanns and Johan Mentink for helpful discussions and for providing us with the s-wave scattering lengths for $f=2$. This work was supported by the DFG, the EU (OLAQUI) and the AFOSR. FG acknowledges support from a Marie-Curie fellowship.

\appendix*
\section*{Coupling strengths}
It is important to notice that for the collision of two spin-1 atoms only one scattering length difference is sufficient to describe the dynamics; for the case of two colliding spin-2 particles there are two relevant scattering length differences. For $f=1$ this is the difference $a_2 - a_0$, and for $f=2$ there are two differences $a_2-a_0$ and $a_4-a_2$. In most of the cases it is, however, more convenient to use derived quantities that directly describe the dynamics for the actual experimental situation.
\subsection*{Rabi dynamics} \label{sec:appendixRabi}
Following the line of argument given in Section \ref{sec:Theory} the parameters $\Omega_{if}$ and $\delta_\mathrm{int}$ of the Rabi model can be calculated. Table \ref{tab:RabiSummary} summarizes the dependency of each relevant parameter in the Rabi model on the specific scattering length. It should be noted that there the scattering lengths $a_0$ and $a_2$ have different values depending on the hyperfine manifold, $f=1$ or $f=2$.

As described in Section \ref{sec:data} we fit the measured oscillation frequency vs magnetic field with only one (two) free parameters for $f=1$ ($f=2$), independent of our uncertainty in optical potential depth. For this, we express all relevant quantities in terms of interaction detunings $\delta_\mathrm{int}$ (see last column in Table \ref{tab:RabiSummary}). 
 
\begin{table*}[htbp]
	\begin{center}
		\begin{tabular}{|c|c|c|c|c|}\hline
			$f$   &Process& Parameter& Corresponding $\Delta a$ & \\ \hline \hline
			                &    & $\Omega_{if}$ & $2 \sqrt{2}/3 (a_2 -  a_0)$ & $2 \sqrt{2} \delta_1$ \\ \cline{3-5}
		  \raisebox{1.5ex}[-1.5ex]{$f=1$} & \raisebox{1.5ex}[-1.5ex]{$\ket{0,0} \leftrightarrow \ket{+1,-1}$} & $\delta_{\mathrm{int}} \equiv \delta_1$    & $(a_2 - a_0)/3$  &  $\delta_1$  \\ \hline 
		        &                                         & $\Omega_{if}$ & $2 \sqrt{2}/35[7(a_2-a_0)+12(a_4-a_2)]$ & $2 \sqrt{2} (\delta_{2,1} + 2 \delta_{2,2})$\\ \cline{3-5} 
		  \raisebox{1.5ex}[-1.5ex]{$f=2$} & \raisebox{1.5ex}[-1.5ex]{$\ket{0,0} \leftrightarrow \ket{+1,-1}$} & $\delta_{\mathrm{int}} \equiv \delta_{2,1}$    & $[7(a_2-a_0)+2(a_4-a_2)]/35$ & $\delta_{2,1}$  \\ \hline
		        &                                         & $\Omega_{if}$ & $2 \sqrt{2}/35 [-7(a_2-a_0)+3(a_4-a_2)]$  & $2 \sqrt{2} (\delta_{2,2} - \delta_{2,1})$ \\ \cline{3-5}
		  \raisebox{1.5ex}[-1.5ex]{$f=2$} &  \raisebox{1.5ex}[-1.5ex]{$\ket{0,0} \leftrightarrow \ket{+2,-2}$}& $\delta_{\mathrm{int}}$    &  $[7(a_2-a_0)+17(a_4-a_2)]/35$ & $(\delta_{2,1} + 3 \delta_{2,2})$\\ \hline
		    		&                                         & $\Omega_{if}$ & $4/35[7(a_2-a_0)+2(a_4-a_2)]$ & $4 \delta_{2,2}$ \\ \cline{3-5}
		  \raisebox{1.5ex}[-1.5ex]{$f=2$} & \raisebox{1.5ex}[-1.5ex]{$\ket{+1,-1}\leftrightarrow \ket{+2,-2}$}& $\delta_{\mathrm{int}}$    & $3/7 (a_4-a_2)  $ & $3 \delta_{2,2}$ \\ \hline
		  			&                                         & $\Omega_{if}$ & $4\sqrt{3}/7(a_4 - a_2)$  & $4 \sqrt{3} \delta_{2,2}$ \\ \cline{3-5}
		  \raisebox{1.5ex}[-1.5ex]{$f=2$} & \raisebox{1.5ex}[-1.5ex]{$\ket{-1,-1} \leftrightarrow \ket{0,-2}$}& $\delta_{\mathrm{int}} \equiv \delta_{2,2}$    & $(a_4-a_2)/7$  & $\delta_{2,2}$\\ \hline
		\end{tabular}
	\end{center}
	\caption{Summary of the collisional coupling strengths and detunings for the experimentally investigated cases as a function of the spin-dependent scattering lengths $a_F$. The last column gives expressions of the various parameters in terms of interaction detunings $\delta_i$ of the measured processes used in the fitting procedure predented in Section \ref{sec:ScatLengthDif}.}
	\label{tab:RabiSummary}
\end{table*}
\bibliographystyle{unsrt}
\bibliography{general}

\begin{thebibliography}{10}

\bibitem{Stenger98}
J.~Stenger, S.~Inouye, D.~M. Stamper-Kurn, H.~J. Miesner, A.~P. Chikkatur, and
  W.~Ketterle.
\newblock Spin domains in ground-state {B}ose-{E}instein condensates.
\newblock {\em Nature}, 396:345, 1998.

\bibitem{Barrett01}
M.~D. Barrett, J.~A. Sauer, and M.~S. Chapman.
\newblock {A}ll-{O}ptical formation of an {A}tomic {B}ose-{E}instein
  {C}ondensate.
\newblock {\em Phys.~Rev.~Lett.}, 87:010404, 2001.

\bibitem{Schmaljohann04}
H.~Schmaljohann, M.~Erhard, J.~Kronj\"ager, M.~Kottke, S.~van Staa,
  L.~Cacciapuoti, J.~J. Arlt, K.~Bongs, and K.~Sengstock.
\newblock Dynamics of ${F}=2$ {S}pinor {B}ose-{E}instein {C}ondensates.
\newblock {\em Phys.~Rev.~Lett.}, 92:040402, 2004.

\bibitem{Chang04}
M.-S. Chang, C.~D. Hamley, M.~D. Barrett, J.~A. Sauer, K.~M. Fortier, W.~Zhang,
  L.~You, and M.~S. Chapman.
\newblock Observation of {S}pinor {D}ynamics in {O}ptically {T}rapped
  $^{87}${R}b {B}ose-{E}instein {C}ondensates.
\newblock {\em Phys.~Rev.~Lett.}, 92:140403, 2004.

\bibitem{Kuwamoto04}
T.~Kuwamoto, T.~Araki, T.~Eno, and T.~Hirano.
\newblock Magnetic field dependence of the dynamics of $^{87}${R}b spin-2
  {B}ose-{E}instein condensates.
\newblock {\em Phys.~Rev.~A}, 69:063604, 2004.

\bibitem{Higbie05}
J.~M. Higbie, L.~E. Sadler, S.~Inouye, A.~P. Chikkatur, S.~R. Leslie, K.~L.
  Moore, V.~Savalli, and D.~M. Stamper-Kurn.
\newblock Direct {N}ondestructive {I}maging of {M}agnetization in a {S}pin-1
  {B}ose-{E}instein {G}as.
\newblock {\em Phys.~Rev.~Lett.}, 95:050401, 2005.

\bibitem{McGuirk02}
J.~M. McGuirk, H.~J. Lewandowski, D.~M. Harber, T.~Nikuni, J.~E. Williams, and
  E.~A. Cornell.
\newblock Spatial {R}esolution of {S}pin {W}aves in an {U}ltracold {G}as.
\newblock {\em Phys.~Rev.~Lett.}, 89:090402, 2002.

\bibitem{Gu04}
Q.~Gu, K.~Bongs, and K.~Sengstock.
\newblock Spin waves in ferromagnetically coupled spinor {B}ose gases.
\newblock {\em Phys.~Rev.~A}, 70:063609, 2004.

\bibitem{Law98}
C.~K. Law, H.~Pu, and N.~P. Bigelow.
\newblock Quantum {S}pins {M}ixing in {S}pinor {B}ose-{E}instein {C}ondensates.
\newblock {\em Phys.~Rev.~Lett.}, 81:5257, 1998.

\bibitem{Pu99}
H.~Pu, C.~K. Law, S.~Raghavan, J.~H. Eberly, and N.~P. Bigelow.
\newblock Spin-mixing dynamics of a spinor {B}ose-{E}instein condensate.
\newblock {\em Phys.~Rev.~A}, 60:1463, 1999.

\bibitem{Lewandowski02}
H.~J. Lewandowski, D.~M. Harber, D.~L. Whitaker, and E.~A. Cornell.
\newblock Observation of {A}nomalous {S}pin-{S}tate {S}egregation in a
  {T}rapped {U}ltracold {V}apor.
\newblock {\em Phys.~Rev.~Lett.}, 88:070403, 2002.

\bibitem{Ho98}
Tin-Lun Ho.
\newblock {S}pinor {B}ose {C}ondensates in {O}ptical {T}raps.
\newblock {\em Phys.~Rev.~Lett.}, 81(4):742, 1998.

\bibitem{Ohmi98}
T.~Ohmi and K.~Machida.
\newblock Bose-einstein condensation with internal degrees of freedom in alkali
  atom gases.
\newblock {\em J.~Phys.~Soc.~Jpn.}, 67:1822, 1998.

\bibitem{Ciobanu00}
C.~V. Ciobanu, S.-K. Yip, and Tin-Lun Ho.
\newblock Phase diagrams of ${F}=2$ {B}ose-{E}instein condensates.
\newblock {\em Phys.~Rev.~A}, 61:033607, 2000.

\bibitem{Koashi00}
M.~Koashi and M.~Ueda.
\newblock Exact {E}igenstates and {M}agnetic {R}esponse of {S}pin-1 and
  {S}pin-2 {B}ose {E}instein {C}ondensates.
\newblock {\em Phys.~Rev.~Lett.}, 84:1066, 2000.

\bibitem{Klausen01}
N.~N. Klausen, J.~L. Bohn, and Ch.~H. Greene.
\newblock Nature of spinor {B}ose-{E}instein condensates in rubidium.
\newblock {\em Phys.~Rev.~A}, 64:053602, 2001.

\bibitem{Dalibard1999a}
J.~Dalibard.
\newblock Collisional dynamics of ultra-cold atomic gases.
\newblock In M.~Inguscio, S.~Stringari, and C.E. Wieman, editors, {\em
  Proceedings of the International School of Physics - Enrico Fermi}, page 321.
  IOS Press, 1999.

\bibitem{Heinzen1999a}
D.~J. Heinzen.
\newblock Ultracold atomic interactions.
\newblock In M.~Inguscio, S.~Stringari, and C.E. Wieman, editors, {\em
  Proceedings of the International School of Physics - Enrico Fermi}, page 351.
  IOS Press, 1999.

\bibitem{Ueda02}
M.~Ueda and M.~Koashi.
\newblock Theory of spin-2 {B}ose-{E}instein condensates: {S}pin correlations,
  magnetic response, and excitation spectra.
\newblock {\em Phys.~Rev.~A}, 65:063602, 2002.

\bibitem{Ho00}
T.-L. Ho and S.~K. Yip.
\newblock Fragmented and single condensate ground states of spin-1 bose gas.
\newblock {\em Phys.~Rev.~Lett.}, 84:4031, 2000.

\bibitem{Pu00}
H.~Pu and P.~Meystre.
\newblock Creating {M}acroscopic {A}tomic {E}instein-{P}odolski-{R}osen
  {S}tates from {B}ose-{E}instein {C}ondensates.
\newblock {\em Phys.~Rev.~Lett.}, 85:3987, 2000.

\bibitem{Duan00}
L.-M. Duan, A.~S{\o}rensen, J.~I. Cirac, and P.~Zoller.
\newblock {S}queezing and {E}ntanglement of {A}tomic {B}eams.
\newblock {\em Phys.~Rev.~Lett.}, 85:3991, 2000.

\bibitem{Mustecaplioglu02}
O.~M{\"u}stecapl{\i}o{\u{g}}lu, M.~Zhang, and L.~You.
\newblock Spin squeezing and entanglement in spinor condensates.
\newblock {\em Phys.~Rev.~A}, 66:033611, 2002.

\bibitem{Demler02}
E.~Demler and F.~Zhou.
\newblock {S}pinor {B}osonic {A}toms in {O}ptical {L}attices: {S}ymmetry
  {B}reaking and {F}ractionalization.
\newblock {\em Phys.~Rev.~Lett.}, 88:163001, 2002.

\bibitem{Yip03}
S.~K. Yip.
\newblock Dimer {S}tate of {S}pin-1 {B}osons in an {O}ptical {L}attice.
\newblock {\em Phys.~Rev.~Lett.}, 90:250402, 2003.

\bibitem{Hou03}
J.-M. Hou and M.-L. Ge.
\newblock Quantum phase transition of spin-2 cold bosons in an optical lattice.
\newblock {\em Phys.~Rev.~A}, 67:063607, 2003.

\bibitem{Svidzinsky03}
A.~A. Svidzinsky and S.~T. Chui.
\newblock Insulator-superfluid transition of spin-1 bosons in an optical
  lattice in magnetic field.
\newblock {\em Phys.~Rev.~A}, 68:043612, 2003.

\bibitem{Imambekov03}
A.~Imambekov, M.~Lukin, and E.~Demler.
\newblock Spin-exchange interaction of spin-one bosons in optical lattices:
  {S}inglet, nematic, and dimerized phases.
\newblock {\em Phys.~Rev.~A}, 68:063602, 2003.

\bibitem{Paredes03}
B.~Paredes and J.~I. Cirac.
\newblock From {C}ooper {P}airs to {L}uttinger {L}iquids with {B}osonic {A}toms
  in {O}ptical {L}attices.
\newblock {\em Phys.~Rev.~Lett.}, 90:150402, 2003.

\bibitem{Snoek04}
M.~Snoek and F.~Zhou.
\newblock Microscopic wave functions of spin-singlet and nematic {M}ott states
  of spin-one bosons in high-dimensional bipartite lattices.
\newblock {\em Phys.~Rev.~A}, 69:094410, 2004.

\bibitem{Jin04}
S.~Jin, J.-M. Hou, B.-H. Xie, L.-J. Tian, and M.-L. Ge.
\newblock Superfluid-{M}ott-insulator transition of spin-2 cold bosons in as
  optical lattice in a magnetic field.
\newblock {\em Phys.~Rev.~A}, 70:023605, 2004.

\bibitem{Tsuchiya04}
S.~Tsuchiya, S.~Kurihara, and T.~Kimura.
\newblock Superfluid-{M}ott insulator transition of spin-1 bosons in an optical
  lattice.
\newblock {\em Phys.~Rev.~A}, 70:043628, 2004.

\bibitem{GarciaRipoll04}
J.~J. Garc{\'i}a-Ripoll, M.~A. Martin-Delgado, and J.~I. Cirac.
\newblock Implementation of {S}pin {H}amiltonians in {O}ptical {L}attices.
\newblock {\em Phys.~Rev.~Lett.}, 93:250405, 2004.

\bibitem{Chang05}
M.-S. Chang, Q.~Qin, W.~Zhang, L.~You, and M.~S. Chapman.
\newblock Coherent spinor dynamics in a spin-1 {B}ose condensate.
\newblock {\em Nature Physics}, 1:111, 2005.

\bibitem{Kronjaeger05}
J.~Kronj{ä}ger, C.~Becker, M.~Brinkmann, R.~Walser, P.~Navez, K.~Bongs, and
  K.~Sengstock.
\newblock Evolution of a spinor condensate: {C}oherent dynamics, dephasing and
  revivals.
\newblock {\em Phys.~Rev.~A}, 72:063619, 2005.

\bibitem{Saito05}
H.~Saito and M.~Ueda.
\newblock Diagnostics for the ground state phase of a spin-2 {B}ose-{E}instein
  condensate.
\newblock {\em Phys.~Rev.~A}, 72:053628, 2005.

\bibitem{Zwerger03}
W.~Zwerger.
\newblock Mott-{H}ubbard transition of cold atoms in optical lattices.
\newblock {\em J.~Opt.~B}, 5:S9, 2003.

\bibitem{MentinkPrivate05}
J.~H. Mentink and S.~J.~J.~M.~F. Kokkelmans.
\newblock Private communication.

\bibitem{Greiner02b}
M.~Greiner, O.~Mandel, T.~W. H{\"a}nsch, and I.~Bloch.
\newblock Collapse and revival of the matter wave field of a {B}ose-{E}instein
  condensate.
\newblock {\em Nature}, 419:51, 2002.

\bibitem{Stamper-Kurn00}
D.~M. Stamper-Kurn and W.~Ketterle.
\newblock In \emph{Coherent Matter Waves}, edited by R.~Kaiser, C.~Westbrook,
  and F.~David (Springer, New York, 2000).
\newblock arXiv: cond-mat/0005001.

\bibitem{vanKempen02}
E.~G.~M. van Kempen, S.~J.~J.~M.~F. Kokkelmans, D.~J. Heinzen, and B.~J.
  Verhaar.
\newblock Interisotope {D}etermination of {U}ltracold {R}ubidium {I}nteractions
  from {T}hree {H}igh-{P}recision {E}xperiments.
\newblock {\em Phys.~Rev.~Lett.}, 88:093201, 2002.

\bibitem{Gerbier05b}
F.~Gerbier, A.~Widera, S.~F\"{o}lling, O.~Mandel, and I.~Bloch.
\newblock Resonant control of spin dynamics in ultracold quantum gases by
  microwave dressing.
\newblock {\em arXiv}, cond-mat:0601151, 2005.
\newblock To be published in \emph{Phys.~Rev.~A}.

\bibitem{Huang05}
H.-J. Huang and G.-M. Zhang.
\newblock Quantum beat phenomenon presence in coherent spin dynamics of spin-2
  {$^{87}$R}b atoms in a deep optical lattice.
\newblock {\em arXiv}, cond-mat:0601188, 2006.

\bibitem{Fewell97}
M.~P. Fewell, B.~W. Shore, and K.~Bergmann.
\newblock {C}oherent {P}opulation {T}ransfer among {T}hree {S}tates: {F}ull
  {A}lgebraic {S}olutions and the {R}elevance of {N}on {A}diabatic {P}rocesses
  to {T}ransfer by {D}elayed {P}ulses.
\newblock {\em Aust.~J.~Phys.}, 50:281--308, 1997.

\bibitem{Jaksch02}
D.~Jaksch, V.~Venturi, J.~I. Cirac, C.~J. Williams, and P.~Zoller.
\newblock Creation of a {M}olecular {C}ondensate by {D}ynamically {M}elting a
  {M}ott {I}nsulator.
\newblock {\em Phys.~Rev.~Lett.}, 89:040402, 2002.

\bibitem{Gericke06}
T.~Gericke, F.~Gerbier, A.~Widera, S.~F{\"o}lling, O.~Mandel, and I.~Bloch.
\newblock Adiabatic loading of a {B}ose-{E}instein condensate in a 3{D} optical
  lattice.
\newblock {\em arXiv}, cond-mat:0603590, 2006.

\bibitem{Jaksch98}
D.~Jaksch, C.~Bruder, J.~I. Cirac, C.W. Gardiner, and P.~Zoller.
\newblock {C}old {B}osonic {A}toms in {O}ptical {L}attices.
\newblock {\em Phys.~Rev.~Lett.}, 81:3108, 1998.

\bibitem{Batrouni02}
G.~G. Batrouni, V.~Rousseau, R.~T. Scalettar, M.~Rigol, A.~Muramatsu, P.~J.~H.
  Denteneer, and M.~Troyer.
\newblock {M}ott {D}omains of {B}osons {C}onfined in {O}ptical {L}attices.
\newblock {\em Phys.~Rev.~Lett.}, 89:117203, 2002.

\bibitem{Gerbier05}
F.~Gerbier, A.~Widera, S.~F\"{o}lling, O.~Mandel, T.~Gericke, and I.~Bloch.
\newblock Phase {C}oherence of an {A}tomic {M}ott {I}nsulator.
\newblock {\em Phys.~Rev.~Lett.}, 95:050404, 2005.

\bibitem{Gerbier05a}
F.~Gerbier, A.~Widera, S.~F\"{o}lling, O.~Mandel, T.~Gericke, and I.~Bloch.
\newblock Interference pattern and visibility of a {M}ott insulator.
\newblock {\em Phys.~Rev.~A}, 72:053606, 2005.

\bibitem{Burt97}
E.~A. Burt, R.~W. Ghrist, C.~J. Myatt, M.~J. Holland, E.~A. Cornell, and C.~E.
  Wieman.
\newblock Coherence, {C}orrelations and {C}ollisions: {W}hat {O}ne {L}earns
  about {B}ose-{E}instein {C}ondensates from {T}heir {D}ecay.
\newblock {\em Phys.~Rev.~Lett.}, 79:337, 1997.

\bibitem{Myatt97}
C.~J. Myatt, E.~A. Burt, R.~W. Ghrist, E.~A. Cornell, and C.~E. Wieman.
\newblock Production of {T}wo {O}verlapping {B}ose-{E}instein {C}ondensates by
  {S}ympathetic {C}ooling.
\newblock {\em Physical Review Letters}, 78:586, 1997.

\bibitem{Widera05}
A.~Widera, F.~Gerbier, S.~F\"olling, T.~Gericke, O.~Mandel, and I.~Bloch.
\newblock Coherent collisional spin dynamics in optical lattices.
\newblock {\em Phys.~Rev.~Lett.}, 95:190405, 2005.

\bibitem{YouPresentation06}
L.~You.
\newblock Oral presentation at the annual meeting of the German Physical
  Society, Frankfurt, 2006.

\bibitem{Yi04}
S.~Yi and L.~You.
\newblock Calibrating {D}ipolar {I}nteraction in an {A}tomic {C}ondensate.
\newblock {\em Phys.~Rev.~Lett.}, 92:193201, 2004.

\bibitem{Jauregui01}
R.~J{\'a}uregui, N.~Poli, G.~Roati, and G.~Modugno.
\newblock Anharmonic parametric excitation in optical lattices.
\newblock {\em Phys.~Rev.~A}, 64:033403, 2001.

\bibitem{GreinerPhD03}
M.~Greiner.
\newblock Ultracold quantum gases in three-dimensional optical lattice
  potentials.
\newblock Dissertation in the {P}hysics department of the
  {L}udwig-{M}aximilians-{U}niversit\"at {M}\"unchen., 2003.
\newblock Online at http://edoc.ub.uni-muenchen.de/archive/00000968/.

\end{thebibliography}
\end{document}